\documentclass{ab2018}
\usepackage{url}
\usepackage{graphicx}
\usepackage{natbib}

\newcommand{\km}{\,\mbox{km s$^{-1}$}}%
\newcommand{\SII}{[S\,II]}%
\newcommand{\OI}{[O\,I]}%
\newcommand{\OIII}{[O\,III]}%
\newcommand{\NII}{[N\,II]}%
\newcommand{\Ha}{H$\alpha$}%
\newcommand{\Hb}{H$\beta$}%

\begin{document} 

\title{Diagnostics of ionized gas in galaxies with the ``BPT--radial velocity dispersion'' relation}

\author{D.V.~Oparin$^{1*}$ and A.V.~Moiseev$^{2***}$}

\institute{$^1$Special Astrophysical Observatory, Russian Academy of Sciences, Nizhnij Arkhyz, 369167  Russia\\
$^2$S2Space Research Institute, Moscow, 117997 Russia
}


 
\titlerunning{Diagnostics of ionized gas in galaxies}

\authorrunning{Oparin \& Moiseev}

\date{March 30, 2018/Revised: June 28, 2018}
\offprints{Dmitry Oparin  \email{doparin2@gmail.com} }


\abstract{
In order to study the state of gas in galaxies, diagrams of the
relation of optical emission line fluxes are used allowing one to
separate main ionization sources: young stars in the H\,II
regions, active galactic nuclei, and shock waves. In the
intermediate cases,  when the contributions of
radiation from OB stars and from shock waves mix, identification
becomes uncertain, and the issue remains unresolved on what
determines the observed state of the diffuse ionized gas (DIG)
including the one on large distances from the galactic plane.
Adding of an extra parameter --- the gas line-of-sight velocity dispersion --- to classical diagnostic diagrams helps to find a
solution. In the present paper, we analyze the observed data for
several nearby galaxies: for UGC\,10043 with the galactic wind,
for the star forming dwarf galaxies VII\,Zw\,403 and Mrk\,35, for the galaxy Arp\,212 with a polar ring. The data on
the velocity dispersion are obtained at the 6-m SAO RAS telescope
with the Fabry-Perot scanning interferometer, the information on
the relation of main emission-line fluxes --- from the published
results of the integral-field spectroscopy (the CALIFA survey and
the MPFS spectrograph). A positive correlation between the radial
velocity dispersion and the contribution of shock excitation to
gas ionization are observed. In particular, in studying Arp\,212,
``BPT--$\sigma$ relation'' allowed us to confirm the assumption on a direct
collision of gaseous clouds on the inclined orbits with the main
disk of the galaxy.
\keywords{galaxies: interstellar medium---galaxies: kinematics and
dynamics---galaxies: star formation}
}

\maketitle

\section{Introduction}

Diagrams of ratios of optical emission-line fluxes are widely used
for diagnostics of gas-ionization sources in galaxies. In the
classic work by \citet*{Bald1981} the two-dimension diagram
of line fluxes of \OIII$\lambda5007$/\Hb\, and
\linebreak\NII$\lambda6583$/\Ha\ was suggested for separation of
objects with different ionization sources. The method became
popular due to the use of measurements of lines bright in the
visible range that are close in wavelengths and, consequently,
with a weak dependence of their intensity ratio on the
interstellar extinction. Later, this method was extended by adding
the relations \SII/\Ha\footnote{Hereinafter, for short we will
designate \OI$\lambda6300$\, as \OI, \OIII$\lambda5007$\, as
\OIII, \NII$\lambda6583$\, as \NII, and
\SII$\lambda6717$+\SII$\lambda6731$\, as \SII.} and
\OI/\Ha~\citep{Veilleux1987,Kewl2001} as
the second parameter. All the mentioned diagrams are frequently
called in the literature the ``BPT diagrams'' after the authors of
the method. Using them, it is possible to distinguish the regions,
where the largest contribution to gas ionization is made by young
massive stars (hereinafter, the H II type) and the regions of
dominating hard ionizing radiation of the active galactic nucleus
(AGN). At the same time, the regions ionized by shock waves, the
asymptotic giant branch (AGB) stars  or nuclei of
galaxies of the LINER type mix in the diagrams
\footnote{Low-Ionization Narrow Emission-line Region in which the
shock ionization of gas can be associated both with a burst of
star formation and with a weak nuclear activity.}. Various variants of
the demarcation lines were
suggested~\citep{Mon2006,Ho2014}, but it is
often problematic to separate the contribution of ionizing sources
with the soft spectrum.

Addition of one more parameter -- the velocity dispersion of the
ionized gas along the line of sight ($\sigma$) -- to the classic
diagnostic diagrams allows one to escape uncertainty in the cases,
when the increase of $\sigma$ indicates the increase of turbulent
velocities of gas beyond the front of a shock wave. However, to
 estimate $\sigma$ reliably, the spectral resolution is necessary
that is noticeably better than that usually required for measuring
the fluxes of radial velocities of separate spectral lines. Thus,
until recently, the dependence of the relation of line fluxes
characterizing the shock ionization from  $\sigma$ was rarely
considered and mainly for the objects with $\sigma>100$--$200\km$
such as galaxies with intense star
formation~\citep{Mon2006,Ho2014}. Such an
approach has not previously been used to study the ionization of
diffuse gas in dwarf galaxies, around separate star-forming
regions, or at some distance from the plane of the galactic disk.

There is a discussion about the sources of ionization of this
diffuse ionized gas (DIG) in galaxies whose role is assigned to an
old stellar population, leakage of Lyman quanta from H\,II
regions, and also possibly to shock fronts caused by star
formation processes \citep[see references and discussion in][]{Jone2017,Egorov2017}. The most
effective methods for studying the extended low-brightness
structures in galaxies are panoramic spectroscopy also called
integral-field, or 3D. In a recent
paper  based on the results of the
SDSS MaNGA  survey by \citep{Zhang2017}, it was concluded that DIG is
associated mainly with the evolved stellar population (AGB stars,
etc.). At the same time, in Section 6.2 of the paper cited, it was
noted that shock waves can be the cause of the observed increase
in the flux ratio of the forbidden and Balmer lines. It is
difficult to verify, since the spectral resolution of the MaNGA
survey is about two times poorer than that required to be able see
the effects of moderate shock waves (with a velocity of less than
500~\km) in the observed kinematics of the ionized gas.
Unfortunately, most of the available observed data on
spectrophotometry and kinematics of the gas of nearby galaxies are
obtained with the spectral resolution $FWHM>5$~\AA~which
corresponds to values greater than 100~\km\ in terms of radial
velocity dispersion or greater than 230 \km\ in terms of the
$FWHM$ in the \Ha line. Observations with such resolution are a
compulsory compromise in the study of low-surface-brightness
objects.

In the SAMI  survey of galaxies  \citep{Ho2014} with the
3D spectroscopy at the 3.9-m Anglo-Australian Telescope (AAT), the
``line ratio--velocity dispersion'' diagrams were built for the
galaxies with active star formation. A positive correlation  of the ionized gas $\sigma$ with a characteristic emission lines
ratios was noticed, which was interpreted as an increase of shock
waves contribution with velocities of about 200-300 \km\
accompanying a burst of star formation. The spectral resolution of
the SAMI survey is greater than that of MaNGA and equals
$R\approx4500$.

A significant limitation of these two most massive today  3D
spectroscopy surveys of galaxies is rather low spatial resolution
(more than 1 kpc). It is related to the fact that the field of
view of integral field unit (IFU) is small and is about
$15\arcsec$ in SAMI~\citep{SAMI2012} and
$12\arcsec$--$32\arcsec$ in SDSS
MaNGA~\citep{MANGA2015}. In these surveys, relatively
distant ($z>0.01$) galaxies are studied. At the same time, the
largest contribution to the kinematics of interstellar medium from
motion due to supernovae and winds of young stars in star-forming
regions is made on considerably smaller spatial scales (from tens
to hundreds of parsec). Consequently, any observed manifestations
of shock fronts in star-forming regions become unevident, when
averaging over a scale of one kpc or more. The examples of
decreasing of peak velocity dispersion of the ionized gas in dwarf
galaxies with   degradation  of a spatial resolution are presented
in~\cite{Mois2012}. \cite{Vasiliev2015} considered the same effect   in simulations of multiple supernova explosions.
Therefore, for observational studies of the relation between an
ionization state of gas and  dispersion of its radial velocities
in galaxies without an active nucleus and with a moderate star
formation rate, 3D spectroscopic data  are required
simultaneously with a considerably high spectral and spatial
resolution.

In this paper, we consider this relation for several nearby
galaxies using a combination of two spectroscopic methods with
similar spatial resolution and quite a large field of view. Velocity 
dispersion map are derived  from the  observations
with a scanning  Fabry-Perot interferometer (FPI) at the 6-m SAO
RAS telescope. Information on the main emission lines ratios are
taken from open data on the integral-field spectroscopy with low
spectral resolution.

In order to show the relation between the velocity dispersion and
the lines ratios characterizing the ionization state, we use
various methods through our paper: coloring in BPT diagrams, the
``$\sigma$--line-flux relation'' diagrams, and
``$\sigma$--distance from the H\,II/AGN demarcation line.'' \textit{As a
general name for the dependencies under study, we use the term
``BPT relation--$\sigma$''}. Classical BPT--diagrams are
two-dimensional plots, where the axes represent relations of line
fluxes. The inclusion of the velocity dispersion in the analysis
is equivalent to the transition to three-dimensional plots, where
a coordinate axis $\sigma$ is added to each diagram. The more
familiar two-dimensional plots given in our paper and in the above
papers are a projection of the BPT--$\sigma$ common relation to
the selected plane.

\begin{table*}[] 
\caption{Characteristics of the galaxies under study and parameters of their observations with different methods}
 \label{tab_1} 
 \begin{tabular}{cccccccccc} 
 \hline
Galaxy & $D$, & $M_B$ &\multicolumn{4}{c}{Integral-field spectroscopy} &\multicolumn{3}{c}{Scanning FPI} \\ 
             & Mpc &         &Instrument &$\Delta\lambda$, \AA &$\delta\lambda$, \AA &$\theta\, ''$ &$\Delta\lambda$ &$\delta\lambda$, \AA &$\theta\,''$ \\ 
\hline
UGC\,10043 &34.9 &$-17.6$ &PPAK &3750--7500 &5--9 &2.7 &\NII &1.7 &1.5 \\
 VII\,Zw\,403~(UGC\,6456) &4.34 &$-13.87$ &MPFS &4250--7200 &8 &2.0 &\Ha &0.8 &2.2 \\ 
Mrk\,35~(NCG\,3353) &15.6 &$-17.75$ &PPAK &3750--7500 &5--9 &2.7 &\Ha &0.8 &2.1 \\ 
Arp\,212~(NGC\,7625) &23.5 &$-18.9$ &PPAK &3750--7500 &5--9 &2.7 &\Ha &0.8 &2.7 \\
\hline
\end{tabular} 
\end{table*}

\section{Spectral data and data reduction}

\subsection{Galaxy Sample}

We considered the sample of of galaxies with data on a ionized gas
state  obtained with two  3D spectroscopy methods. When compiling
the sample, we prepared a list of nearby galaxies for which, based
on observations with the scanning FPI at the 6-m telescope of SAO
RAS, the fields of velocity dispersion of ionized gas in the \Ha\,
or \NII\ emission lines were constructed. In total, this is about
60 objects that were observed in 2002--2015; most of them are
presented in \cite{Mois2015}. For each galaxy in
the list, we checked the presence of data cubes in open sources
obtained with the integral-field spectroscopy. For three galaxies:
Arp\,212, Mrk\,35, and UGC\,10043 such data were obtained within
the framework of the CALIFA survey \citep[The Calar-Alto Legacy Integral
Field Area,][]{Sanc2012}. We used the third data
release of CALIFA; the spectra are available on the project
website\footnote{\url{http://califa.caha.es/}}. Let us notice that
UGC 10043 was observed with the 6-m telescope specially on request
of the CALIFA team. The results were presented in our
collaborative paper~\citep{Loko2017} and triggered our
interest for further study of the BPT--$\sigma$ relation.
Nevertheless, to keep homogeneity, we repeated the analysis of the
UGC 10043 data in this paper using the same methods as for other
galaxies. For the galaxy  VII\,Zw\,403, there was a data cube
obtained by combining several fields of the MPFS spectrograph in
observations at the 6-m telescope of SAO RAS and published by \citet{Arc2007}.

\begin{figure*}[] 
\centerline{\includegraphics[scale=0.8]{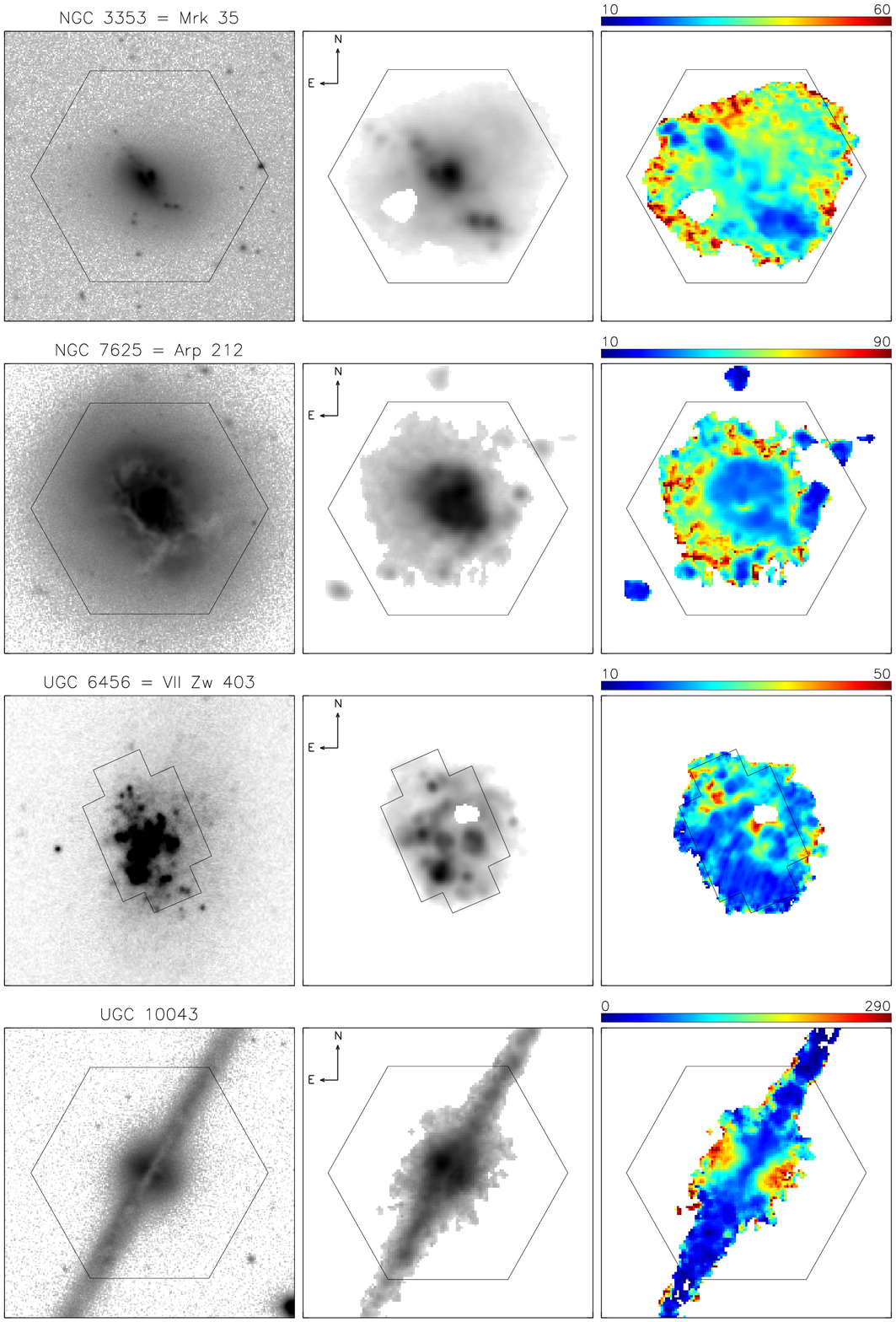}}
\caption{The images of the studied galaxies. The left-hand column
shows the SDSS  $r$-filter images
from~\cite{Knapen2014}, for UGC\,6456 the image in the
$R$ filter is given from~\cite{GildePaz2003}. The
middle column gives the images in the [N\,II]$\lambda6583$ line
(UGC\,10043) or \Ha\ line (for others) from the observed data with
the FPI at the 6-m BTA telescope. The right-hand column shows the
velocity dispersion of the ionized gas; the scale in \km. Image
size is $90''\times90''$. The arrangement of the fields of view of
integral-field spectrographs is shown: the CALIFA survey and the
MPFS mosaic (for UGC 6456). The velocity dispersion fields are
taken from ~\cite{Mois2015}
and~\cite{Loko2017} (for UGC\,10043) without
correction for the thermal line broadening.}
\label{fig_1} 
\end{figure*}

Table~\ref{tab_1} briefly presents the objects under study
(accepted distance $D$ and absolute magnitude $M_B$ based on the
NED data) and on the data used (observation instruments,
$\Delta\lambda$--spectral range or a selected line,
$\delta\lambda$--spectral resolution in terms of $FWHM$,
$\theta\arcsec$--angular resolution). Figure~\ref{fig_1} shows the
images of the sample galaxies  in the $r$ filter and in the
emission lines, and velocity dispersion fields $\sigma$ of the
ionized gas constructed from the scanning FPI data. It also shows the field of view of the spectrographs used.

\subsection{Low-Resolution Spectra}

For UGC\,10043, Mrk\,35, and Arp\,212, we used the CALIFA data
obtained at the 3.5-m telescope of the Calar Alto observatory in
the mode of the integral-field spectroscopy of the PPAK wide
field~\citep{Kelz2006} of the PMAS spectrograph~\citep{Roth2005}.
The array of PPAK optical fibers comprises 331 elements of the
$2\farcs7$ diameter collected in the $74''\times64''$ hexagonal
field. We used the cubes obtained in the low-resolution mode
covering the entire visible range (grating V500, $R\sim850$). The
reduces data are presented in the form of cubes extrapolated from
a hexagonal grid to a square grid with a spatial element size
(spaxel) of $1''$.

The galaxy VII\,Zw\,403 was observed with the MPFS multislit
spectrograph~\citep{Afan2001} at the SAO RAS 6-m  telescope.
An array of square lenses combined with fiber optics provided a
field of view of $16\times16$ elements with a scale of $1''$ per
lens. The data cube presented in   
\cite{Arc2007}  is a mosaic of the size of
\mbox{$49''\times31''$} comprising seven MPFS fields. The spectral
range was 4250--7200~\AA; the resolution was 8~\AA.

For the analysis, we used the data cubes with the $2\times2$
binning   to a scale of $2''$ per element (see next
Section~\ref{sec_kinem}). Approximation of the lines
in the spectra was carried out by the one-component Gaussian
function. The line fluxes ratios were measured only from the
spectra in which $S/N>2$ for each emission line. The local
continuum level near each line was taken into account.

\subsection{Kinematics of the Ionized Gas}

\label{sec_kinem} The archival observations with the
scanning Fabry-Perot interferometer installed at the
SCORPIO~\citep{Afan2005} and
SCORPIO-2~\citep{Afan2011}  focal reducers in the
primary focus of the  6-m  telescope were used to create the
velocity dispersion maps. The emission line (H$\alpha$ or
[N\,II]$\lambda6583$) was selected with the  a narrow-band
filter with a bandwidth $\sim20$\AA. UGC\,10043 was
observed with the IFP186 interferometer providing the spectral
resolution $FWHM=1.7$~\AA. In the study of other galaxies, IFP501
with the resolution $FWHM=0.8$~\AA\ was used. In observations of
VII\,Zw\,403, the image scale was $0\farcs56$~px$^{-1}$ with the
field of view of $4\farcm8$, for other galaxies:
$0\farcs7$~px$^{-1}$ and $6\farcm1$, respectively. The result of
the reduction of the set of interferograms was a data cube, where
each pixel contained the spectrum of the selected emission line
consisting of 36--40 channels. The details of data reduction and
observational logs were published earlier (see references in
Section~\ref{sec_galaxies}). Since the instrumental
contour of the interferometer is well described with the Lorentz
profile, the observed profiles of emission lines were approximated
by the Voigt function --- the convolution of the Gauss and Lorentz
functions~\citep{MoiseevEgorov2008}. It is assumed that
the initial (without any instrumental broadening) profile of an
emission line is satisfactorily described by a Gaussian, which is
a good approximation for observations of H II regions, with the
exception of individual peculiar cases (expanding envelopes,
neighborhood of Wolf-Rayet stars, etc.
\citep[see ][]{Mois2012,Egorov2017}). Based on
the results of approximation, we built the monochromatic images in
this line, the distribution of radial velocities of the ionized
gas, and the radial velocity dispersion maps free from the
instrumental broadening of the spectral line
profile~\citep{MoiseevEgorov2008}.

The $\sigma$ maps in Fig.~\ref{fig_1} are shown with
the original sampling of the FPI images ($0\farcs6$--$0\farcs7$)
which is better than in the data used in the integral-field
spectroscopy ($1''$/spaxel), while the angular resolution $\theta$
of both data sets is similar (see Table~\ref{tab_1}).
To account for this effect, first we interpolated the FPI cube   to a coarser grid corresponding to the  CALIFA or MPFS
data. The accuracy of coincidence of both data sets was controlled
from the images in the emission lines and continuum and was better
than \mbox{$0\farcs2$--$0\farcs5$}. In order to detect the
emission lines in the low-surface-brightness regions in the
optimum way, we performed the pixel binning of \mbox {$2\times2$}
for both combined data sets. This procedure also reduced possible
errors of a small difference between the angular resolution of the
FPI data and integral-field spectroscopy. Therefore, all the
measurements presented below are performed from the cubes with the
$2''$ element size. After combining and binning in the FPI cubes,
we built the maps $\sigma$. In the next section, these maps were
used for direct per-pixel comparison with the low-resolution
spectra.

We used masking to highlight the points with the ratio \mbox {$S/N
\ge3$} in the maps $\sigma$. Let us note that, as distinct from
 \cite{Mois2015}, we did not correct the velocity dispersion
maps for thermal line broadening.

\begin{figure*}[] 
 \centerline{\includegraphics[scale=0.5]{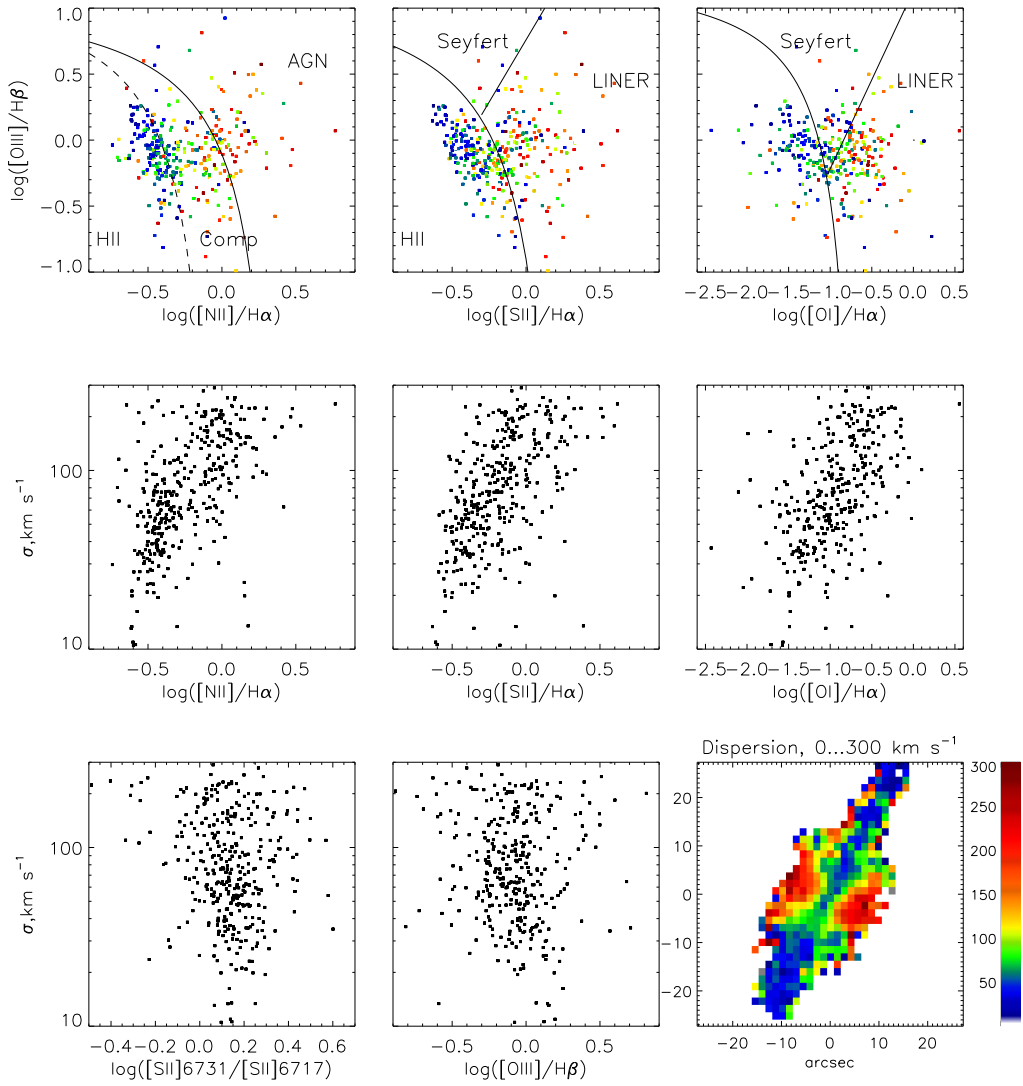}}
\caption{Upper row: BPT diagrams for UGC\,10043. The ionized gas
line-of-sight velocity dispersion in the given pixel according to
the map shown bottom right is colored. The lines separating the
H\,II regions, the objects with the combined ionization type
active Seyfert galaxies, and   LINER  are taken from
\cite{Kewl2006}. The other diagrams: the
dependence between the velocity dispersion and the emission lines
ratios.} \label{diag1}
\end{figure*}

\begin{figure*}[] 
 \centerline{\includegraphics[scale=0.5]{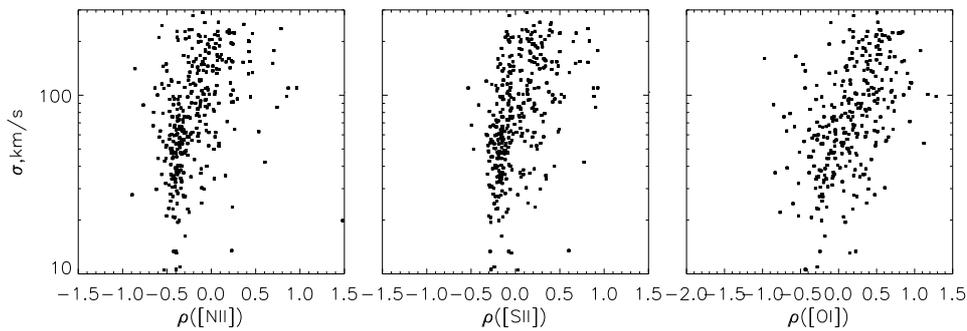}}
\caption{UGC\,10043. Dependence of $\sigma$ on the distance
of the point to the demarcation curve in the BPT diagram
separating the H II regions and regions with other ionization
mechanisms~\citep[accroding][]{Kewl2001}. } \label{diag2} \end{figure*}

\section{BPT--$\sigma$ diagrams}
\label{sec_galaxies}
\subsection{Galactic Wind in UGC\,10043}
UGC\,10043 is an edge-on spiral galaxy. Observations in the
H$\alpha$ and [N\,II] lines carried out with the
HST~\citep{Matt2004} have shown signs of star formation
in the galactic core, as well as an extended emission structure
that is perpendicular to the disk and is the result of the
galactic wind influence. \cite{Loko2017} presented
diagnostic diagrams for the central part of the galaxy  
according to the CALIFA survey. Some of the points on
the diagram relating to the central region of star formation
turned out to be located in the region characteristic of
photoionization by young stars, while others fell into the region
typical of shock excitation. Within the framework of a shock
excitation model, a wind velocity was constrained: no greater than
400~\km. Analyzing the gas velocity field in the \NII\ line
constructed with a scanning FPI allowed us to obtain a more strict
limitation on the galactic wind velocity: less than 250 \km in
accordance with the gas shock excitation model. In the same
paper~\cite{Loko2017}, it has been shown that there is
a distinct relation \mbox{BPT--$\sigma$}  in the wind nebula of
UGC\,10043.

As it is shown in the diagnostic diagrams presented in
Fig.~\ref{diag1} (the upper row), the regions with
shock excitation of emission lines in the wind nebula are
characterized by a higher velocity dispersion as compared to the
regions dominated by photoionization. In this case, there is a
positive correlation between the relations of line fluxes of
[S\,II] to H$\alpha$, of [N\,II] to H$\alpha$, of [O\,I] to
H$\alpha$ and $\sigma$ (see Fig.~\ref{diag1}). At the
same time, negative correlation is observed between the ratio of
the sulfur doublet lines ([S\,II]6731/[S\,II]6717) and $\sigma$.
This means that a higher velocity dispersion is characteristic of
the diffuse gas with a lower electron density $n_e$.

We tried to quantify the BPT--$\sigma$ relation. For each point in
the diagrams of the lines ratios, it is possible to determine the
minimum distance to the curve that bounds the H II-type ionization
region from \cite{Kewl2001} (in the case of the
\mbox{[N\,II]/\Ha--[O\,III]/\Hb} diagram, this is the boundary
between the Comp and AGN regions in our figures). We marked this
distance as $\rho$ and determined it so that negative values of
$\rho$ corresponded to the shift of the points from the
demarcation line to the side corresponding to photoionization by
young stars, and positive--towards other ionization mechanisms.
Figure~\ref{diag2} shows the examples of relations
involved this parameter. For brevity and convenience of reading,
we have designated the value $\rho$ for the
\NII/\Ha--\OIII/\Hb\,diagrams as $\rho([{\rm N\,II}])$, for the
\SII/\Ha--\OIII/\Hb\,\,diagrams as $\rho([{\rm S\,II}])$, and for
the \OI/\Ha--\OIII/\Hb\,~\,diagrams as $\rho([{\rm O\,I}])$. It
can be seen that in all the cases presented, the increase in the
velocity dispersion along the line of sight correlates with the
distance from the region characteristic of ionization by young
stars in the BPT diagram.

\begin{figure*}[]
 \centerline{ \includegraphics[scale=0.5]{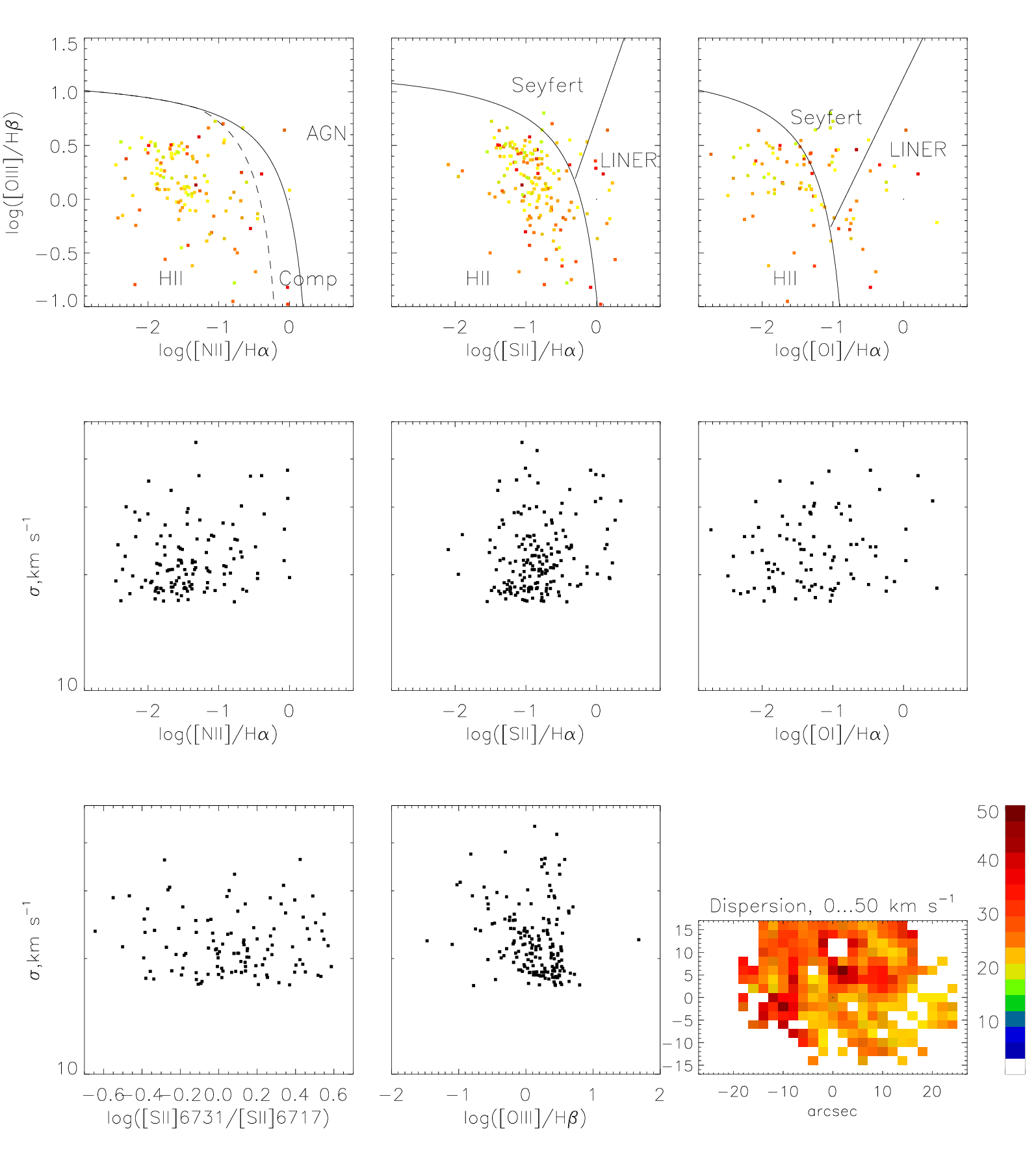} }
\caption{Same as in Fig.~\ref{diag1}, for
VII\,Zw\,403.} \label{diag3} \end{figure*}

\begin{figure*}[]
 \centerline{\includegraphics[scale=0.5]{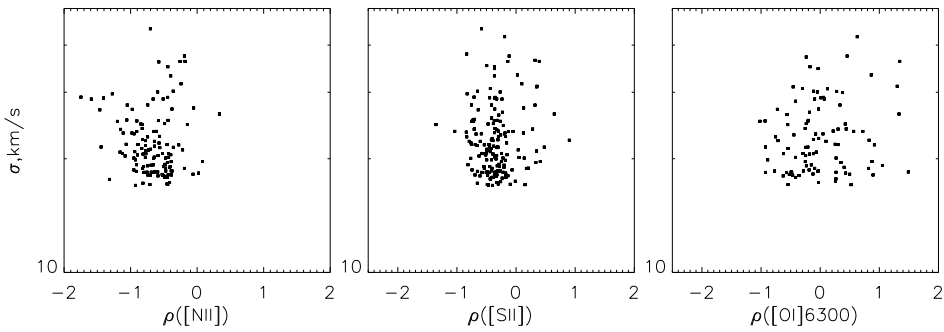} }
\caption{Diagrams similar to those in Fig.~\ref{diag2}, for
VII\,Zw\,403.} \label{diag7a} \end{figure*}

\begin{figure*}[]
 \centerline{ \includegraphics[scale=0.5]{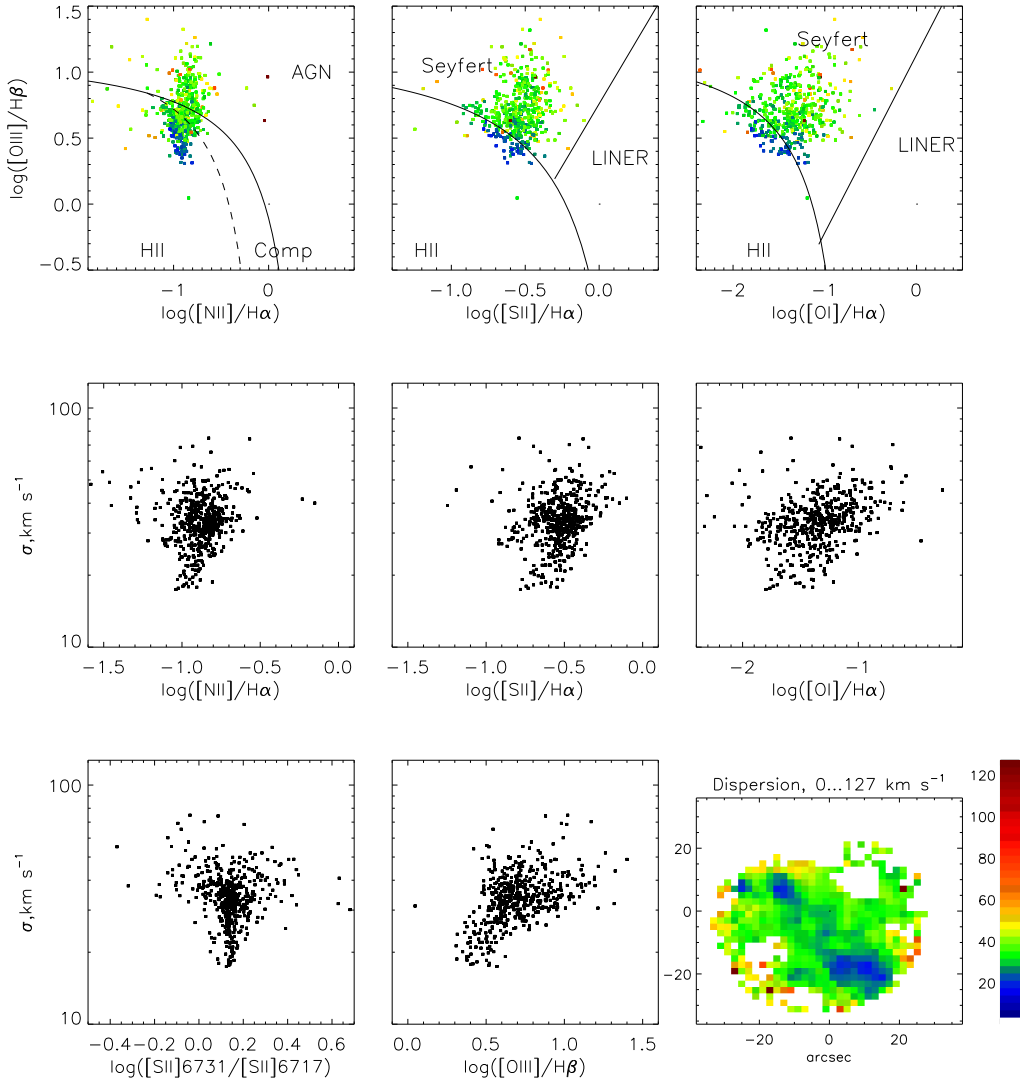}}
\caption{Same as in Fig.~\ref{diag1}, for Mrk\,35.}
\label{diag6} \end{figure*}

\begin{figure*}[] 
 \centerline{\includegraphics[scale=0.5]{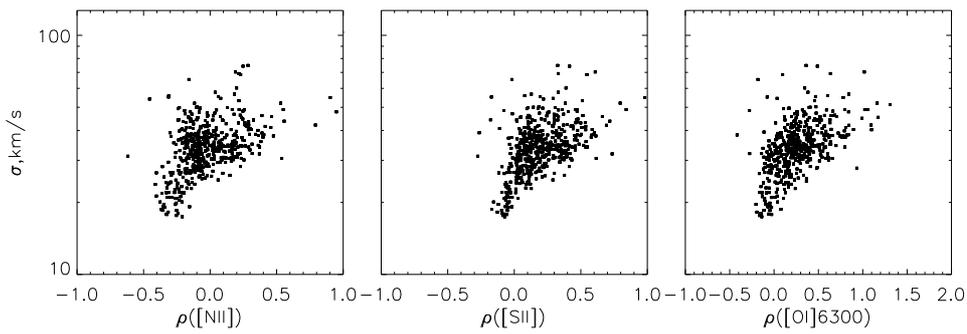} }
\caption{Diagrams similar to those Fig.~\ref{diag2}, for Mrk\,35.}
\label{diag7} \end{figure*}

\begin{figure*}[] 
 \centerline{ \includegraphics[scale=0.5]{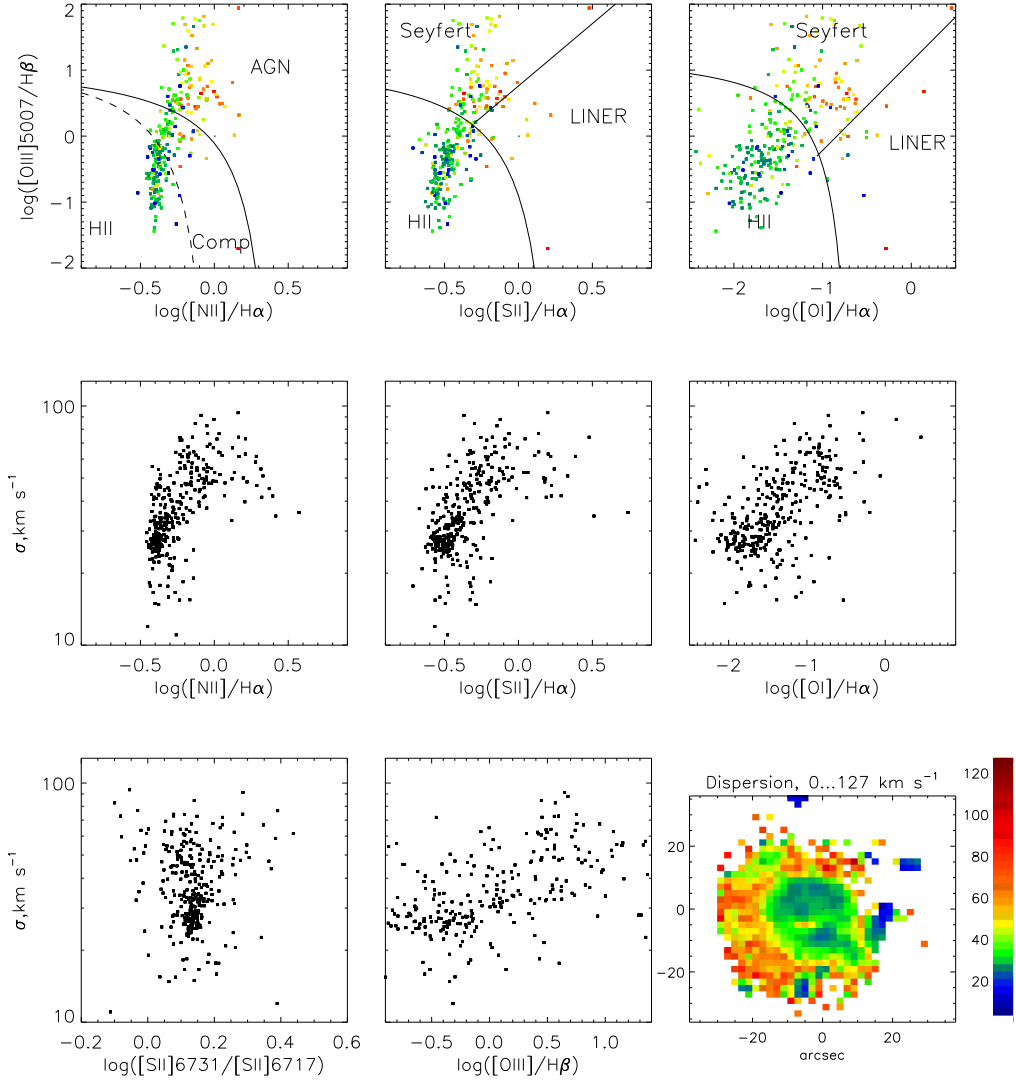} }
\caption{Same as in Fig.~\ref{diag1}, for Arp\,212.}
\label{diag4} \end{figure*}

\begin{figure*}[] 
\centerline{\includegraphics[scale=0.5]{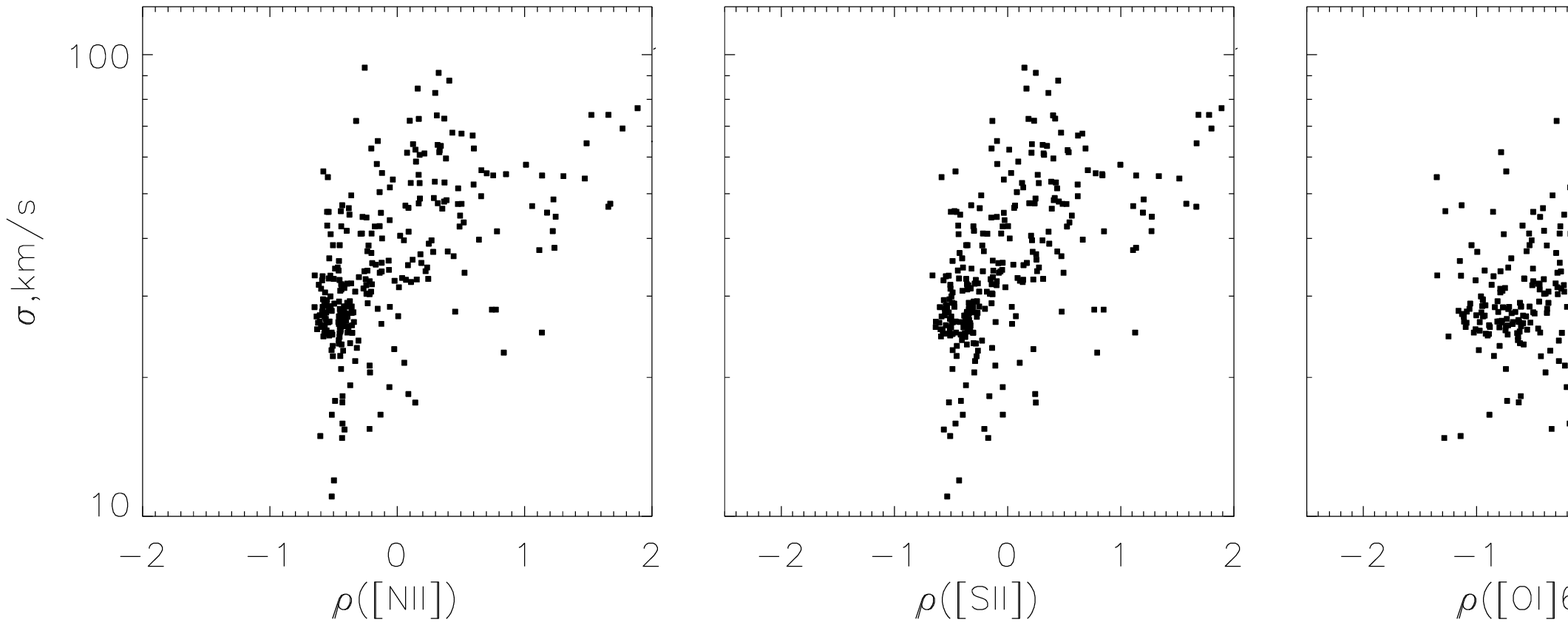}}
\caption{Diagrams similar to those Fig.~\ref{diag2}, for Arp\,212.}
\label{diag5} \end{figure*}

\subsection{Mrk\,35 and VII\,Zw\,403---Dwarf Galaxies with a Burst of Star Formation}
\label{sec_mrk35}

VII\,Zw\,403 is one of the nearest blue compact dwarf galaxies
with several episodes of recent star formation. The current
outburst is located in the central kpc, where several compact
 OB-stars associations are identified and the associated H II
shells that are immersed in the diffuse ionized gas 
\citep[see][ and references therein]{Egor2011}. The fields of
velocities and velocity dispersions of the ionized gas in this
galaxy were previously considered in 
\citet{Loz2006,Mois2012,Mois2015},
where a sufficiently quiet kinematics of the gas with a low level
of peculiar velocities was noticed. The value $\sigma$ is in the
range of 15--40~\km. In the BPT diagrams, most points
are located in the region of photoionization (see
Fig.~\ref{diag3}). A certain number of points with
higher dispersion are found near the separation curve. Along with
this, the expanding H II shells associated with bright star
formation regions are characterized by a smaller value of
$\sigma\sim20$\km. One of the regions with a higher $\sigma$ is
located between these two shells. Others are located on the
periphery of the ionized gas disk~\citep{Mois2012}. In
the ``lines ratio--velocity dispersion'' diagrams, there are no
noticeably significant correlations
(Fig.~\ref{diag3}). Therefore, we can conclude that
the contribution of shock excitation to gas ionization in this
galaxy is negligible and even at the boundaries of the expanding
shells it is noticeably inferior to photoionization (the H II
type). This is also indicated by the absence of significant
correlation between $\rho$ and $\sigma$ in
Fig.~\ref{diag7a}.

Mrk\,35~is another example of a blue compact galaxy. The ongoing
star formation here is concentrated in several bright compact
regions. Star-forming regions near the optical center of the
galaxy form a bar-like structure, where the population of
Wolf--Rayet stars is observed~\citep{Cairos2007}. The
radial velocity dispersion of the ionized gas in the galaxy
reaches about 70~\km, whereas in the central regions it lies
within the range of 20--35~\km. The highest dispersion of radial
velocities is observed in the gas located between three central
regions of star formation. In the ``arms'', the dispersion is
several times lower in comparison with the central regions; and as
a whole does not exceed 20~\km. In the BPT diagrams (see
Fig.~\ref{diag6}), the points corresponding to the
regions with the ongoing star formation are located in the region
of photoionization. The outer parts of the galaxy, characterized
by low surface brightness and high dispersion of radial
velocities, appear near the separation curves which suggests a
certain contribution of shock waves to the gas ionization in these
regions. As well as in UGC\,10043, the sulfur lines ratio in
Fig.~\ref{diag6} demonstrates the anticorrelation.
The $\sigma$--$\rho$ diagrams show a positive correlation between
the distance to the model curve and the velocity dispersion
(Fig.~\ref{diag7}).

\subsection{Arp\,212 -- a Polar Ring Galaxy}

Arp\,212~is a peculiar galaxy in which two rotating gas subsystems
that are kinematically different have been discovered: an internal
disk of the 3.5-kpc size and outer H II regions whose orbits are
inclined at a significant angle to the stellar
disk~\citep{Mois2008}. The observed picture was explained in the
assumption that the gas (mostly neutral) in the outer regions of
the galaxy is located in a wide ring of a diameter of about 20 kpc
rotating in the plane almost orthogonal to the disk. As the radii
of the gaseous-cloud orbits decrease, their inclination angle
decreases too; and at a radius of 2--3 kpc, the gas from the ring
begins to fall out onto the plane of the galaxy inducing a burst
of star formation. It is the region with the highest observed
velocity dispersion   reaching 80--100~\km (see
Fig.~\ref{diag4}). The points belonging to this collision region
of the gas subsystems are shifted in the BPT diagrams (Fig.
\ref{diag4}) from the regions dominated by photoionization towards
the dominance of shock ionization. At the same time,
photoionization clearly dominates in the central region of the
galaxy.

As well as in UGC\,10043 and Mrk\,35, there is a positive
correlation between $\sigma$ and $\rho$ for all the BPT diagrams
(Fig.~\ref{diag5}). It is important to notice that in
all three galaxies this dependence can be observed for the
velocity dispersion above 30--40 \km and practically disappears
for smaller $\sigma$. In other words, the correlation between
$\sigma$ and $\rho$ manifests itself in the presence of shock
excitation in the diffuse gas (DIG) and disappears in the H II
regions characterized by a low level of turbulent motions. This
can also be confirmed by the absence of distinct \mbox
{$\sigma$--$\rho$} correlations in the galaxy VII\,Zw\,403, where
in all the points \mbox {$\sigma<40$}~\km.

As distinct from UGC 10043 and Mrk 35, the relation of sulfur
lines ratio in Arp 212 does not show any pronounced dependence on
$\sigma$ which agrees with the assumption that high velocity
dispersion is observed not only in DIG with low electron density
but also in a denser medium of colliding gaseous clouds.

\section{Conclusion}

For an observational study of the relation between turbulent
motions of the ionized gas in nearby galaxies and the state of its
ionization, it is required to have panoramic spectroscopy data
together with a large field of view and quite high spectral
resolution. Since it is necessary to observe the
low-surface-brightness region with an angular resolution of about
$1''$, then an optical telescope of a large (\mbox {$D>3$--$5$}~m)
diameter is needed. All these requirements are implemented
together probably only in the unique MUSE instrument at the 8-m
VLT telescope~\citep{MUSE2010}.

Our idea is to combine the ionization ionized gas velocity
dispersion maps obtained in the observations with the scanning FPI
and the panoramic spectrophotometry data for
low-spectral-resolution galaxies. The observed line-of-sight
velocity dispersion characterizing the turbulent motions of the
ionized gas can be due to various causes such as virial motions in
the galaxy's gravitational potential, the effect of expanding
shells on the gas, or, more generally, energy injected into the
interstellar medium by star-forming
processes~\citep[see discussion and references in][]{Mois2015,Krum2016}.
Various factors influence the value of line flux relations with
different excitation mechanisms. Observational information fusion makes it
possible, in certain cases, to draw unambiguous conclusions about
contribution of shock waves to the gas ionization in
low-surface-brightness regions. From the lines flux ratio in the
conical nebula in UGC 10043 only, it cannot be definitely
concluded what leads the growth of the relative intensity of the
forbidden lines: ionization by shock waves from the central burst
of star formation or the old stellar population of the thick disk,
in which it is located. Additional information on the gas
kinematics allows   to say that there is a galactic wind. For Arp
212, our approach allowed  to confirm the previously assumption
in \cite{Mois2008} on the direct collision of gaseous
clouds on inclined orbits with the main disk of the galaxy
generating shock fronts.

Thus, the use of the BPT--$\sigma$ diagram together with the
classical diagnostic methods based on lines ratios helps us better
understand of ionization of the galactic interstellar medium in
each specific case. The only galaxy in which we did not find a
correlation between $\sigma$ and characteristic line flux
relations (or $\rho$ parameter)~is VII\,Zw\,403. The ongoing
star-formation rate here is the lowest in our sample 
\citep[$\sim0.015\,M_{\odot}$\,yr$^{-1}$,][]{Loz2006}.
Apparently, for this reason, the contribution of shock waves to
gas ionization is practically invisible.

We plan to conduct further expansion of the sample of the objects
under study in two ways. The first is new observations with a high
spectral resolution of galaxies, for which there are the CALIFA
survey data already, with the scanning FPI. The second is the
creation of images in the emission lines of galaxies, for which we
already have maps of the velocity dispersion of the ionized gas.
Here it is proposed to use a tunable--filter photometer,
the first observations with which are already being conducted by
our
team\footnote{\url{https://www.sao.ru/Doc-en/Events/2017/Moiseev/moiseev_eng.html }}.

\begin{acknowledgements}
The work was supported by the Russian Science Foundation (project
No. 17-12-01335 ``Ionized gas in galactic disks and beyond the
optical radius.'' The paper used the survey data provided by the
Calar Alto Legacy Integral Field Area (CALIFA) survey
(\mbox{\url{ http://califa.caha.es/}}) based on the observations
collected at the Centro Astron{\'o}mico Hispano Alem{\'a}n (CAHA)
at Calar Alto operated jointly with the  Max-Planck-Institut
f{\"u}r Astronomie and the Instituto de Astrof{\'i}sica de
Andaluc{\'i}a (CSIC). This research has made use of the NASA/IPAC
Extragalactic Database (NED) which is operated by the Jet
Propulsion Laboratory, California Institute of Technology, under
contract with the National Aeronautics and Space Administration.
The authors are grateful to Alexandrina Smirnova and the reviewer
for constructive comments.
\end{acknowledgements}


\end{document}